\documentclass[hidelinks]{article}

\usepackage{arxiv}

\usepackage{graphicx}
\usepackage{amsfonts,amsthm,amssymb,amsmath,amsopn,float}
\usepackage{xifthen}
\usepackage{graphicx}
\usepackage{epstopdf}
\usepackage{multicol}
\usepackage{todonotes}
\usepackage{hyperref}

\usepackage{lmodern}

\usepackage{xcolor}

\usepackage[utf8]{inputenc}

\usepackage{enumitem}

\usepackage{comment}

\definecolor{darkgreen}{rgb}{0.0, 0.5, 0.0}
\definecolor{amaranth}{rgb}{0.9, 0.17, 0.31}
\definecolor{azure}{rgb}{0.0, 0.5, 1.0}

\usepackage{listings}

\usepackage{tikz,pgfplots}
\pgfplotsset{compat=1.16}
\usepackage[ruled]{algorithm2e}
\usepackage{algcompatible}
\algblockdefx[Hloop]{Hloop}{EndHloop}[1][]{\textbf{hpx::parallel::for$\_$loop} #1}{\textbf{end parallel for}}
\algblockdefx[Func]{Func}{EndFunc}[1][]{\textbf{Function }}{\textbf{end Function}}

\SetAlFnt{\small\sffamily}

\algrenewcommand\algorithmiccomment[2][\footnotesize]{{#1\hfill\(\triangleright\) #2}}

\makeatletter
\newcommand{\linebreakand}{%
  \end{@IEEEauthorhalign}
  \hfill\mbox{}\par
  \mbox{}\hfill\begin{@IEEEauthorhalign}
}
\makeatother

\newcommand{\R}{\mathbf{R}}  % The real numbers.
\newcommand{\Z}{\mathbf{Z}}  % The integer numbers.
\newcommand{\sref}[2]{\hyperref[#2]{#1 \ref*{#2}}}

\newcommand{\msd}{\,\mathrm{d}}

\newcommand{\orcid}[1]{\href{https://orcid.org/#1}{\includegraphics[height=10pt]{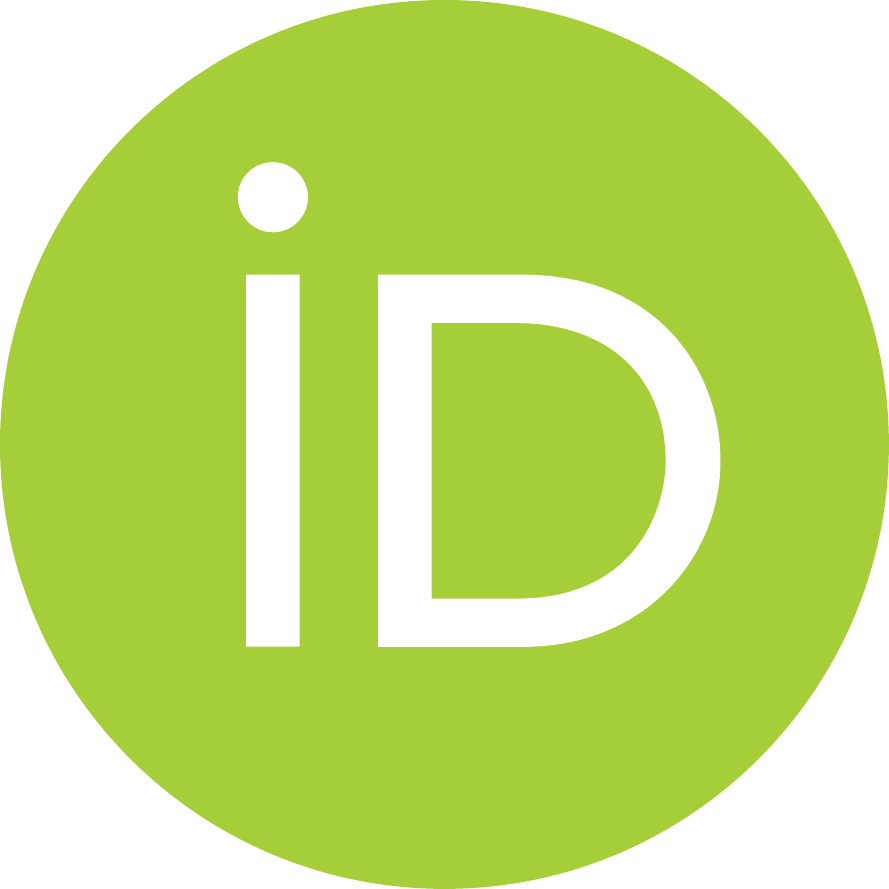}}}

\definecolor{commt}{rgb}{0.2, 0.5, 0.2}
\newcommand{\codecomment}[1]{{\small \textcolor{commt}{$\triangleright$ #1}}}

\newcommand{\SD}{\mathrm{SD}}
\newcommand{\SP}{\mathrm{SP}}
\newcommand{\DP}{\mathrm{DP}}

\begin{document}
%\bstctlcite{IEEEexample:BSTcontrol}
\title{Load balancing for distributed nonlocal models within asynchronous many-task systems}

%Domain decomposition, load balancing, and massively parallel solvers for the class of nonlocal models

\hypersetup{
pdftitle={Load balancing for distributed nonlocal models within asynchronous many-task systems},
pdfsubject={cs.DC},
pdfauthor={Pranav Gadikar, Patrick Diehl, Prashant K. Jha},
pdfkeywords={Nonlocal Computational Models, Load Balancing, AMT, HPX, Parallel and Distributed Computing}
}

\author{
Pranav Gadikar \\
\textit{Computer Science Department}\\ 
\textit{Indian Institute of Technology Madras, Chennai, India}\\
cs16b115@smail.iitm.ac.in
\\
\vspace{0.25cm} 
\\
Patrick Diehl\orcid{0000-0003-3922-8419} \\
\textit{Center for Computation and Technology} \\
\textit{Louisiana State University, Baton Rouge, USA}\\
patrickdiehl@lsu.edu
\\
\vspace{0.25cm} 
\\
Prashant K. Jha\orcid{0000-0003-2158-364X} \\
\textit{Oden Institute for Computational Engineering and Sciences} \\
\textit{The University of Texas at Austin, Austin, USA}\\ 
pjha@utexas.edu
}

\maketitle

\begin{abstract}
In this work, we consider the challenges of developing a distributed solver for models based on nonlocal interactions. In nonlocal models, in contrast to the local model, such as the wave and heat partial differential equation, the material interacts with neighboring points on a larger-length scale compared to the mesh discretization. In developing a fully distributed solver, the interaction over a length scale greater than mesh size introduces additional data dependencies among the compute nodes and communication bottleneck. In this work, we carefully look at these challenges in the context of nonlocal models; to keep the presentation specific to the computational issues, we consider a nonlocal heat equation in a 2d setting. In particular, the distributed framework we propose pays greater attention to the bottleneck of data communication and the dynamic balancing of loads among nodes with varying compute capacity. For load balancing, we propose a novel framework that assesses the compute capacity of nodes and dynamically balances the load so that the idle time among nodes is minimal. Our framework relies heavily on HPX library, an asynchronous many-task run time system. We present several results demonstrating the effectiveness of the proposed framework. 
\end{abstract}


\newpage

%%%%%%%%%%%%%%%%%%%%%%%%%%%%%%%%%%
\section{Introduction}
%%%%%%%%%%%%%%%%%%%%%%%%%%%%%%%%%%
Nonlocal models are seen in various fields; peridynamics for modeling fracture and damage in solid media~\cite{diehl2019review, Jha2020peri}, nonlocal heat and diffusion equation~\cite{burch2011classical}, cell-cell adhesion in computational biology~\cite{armstrong2006continuum, engwer2017structured}, and recently application of peridynamics to granular media~\cite{jha2020peridynamics}. Unlike the models based on the local interaction, for instance, wave and heat partial differential equation, the nonlocal models include the interaction of material points over a finite length scale; as a result, after the spatial discretization of the model, the discrete points interact over a length scale that is larger than the mesh size (minimum spacing between the discrete points).
% removed ``, the so-called ghost zone." as it does not fit well here
In contrast, in the discretization of wave or heat equations, the interaction is local, \emph{i.e.}, the discrete point only interacts with the nearest neighbor points. The larger interaction length scale introduces a major challenge in implementing a \textit{distributed solver}; a node has stronger data dependency with the neighboring node which means that the message size to exchange the ghost zones increases. Thus, we need to overlap communication with computation to address this challenge. We carefully look at this issue in the context of the nonlocal heat equation, which is a simple nonlocal model, but still general enough that framework developed in this work can be applied to more complex models such as peridynamics~\cite{diehl2019review, Jha2020peri}. We investigate the following: (\textit{i}) \textit{Minimizing data exchange} -- here we use the METIS library~\cite{metis_lib} to generate optimal partitions, with a minimum data exchange between nodes. (\textit{ii}) \textit{Hiding data exchange time} -- the partitions are divided such that we have independent sub-partitions with sub-partitions depending on the ghost zones on a different node. Computations on the independent partitions (partitions which do not depend on the data of other compute nodes) are performed asynchronously while waiting for the data from the neighboring nodes for computations on the dependent partitions. (\textit{iii}) \textit{Load balancing} -- the partitions are redistributed among the nodes to minimize the waiting time for the faster nodes. The shape of the partitioning obtained by the METIS library is preserved to the maximum possible extent, in order to reduce the data dependencies.

For the  asynchronous function computation and synchronization, we rely heavily on the HPX~\cite{Kaiser2020} library; HPX is the C++ standard library for parallelism and concurrency for high-level programming abstractions and provides wait-free asynchronous execution and futurization for synchronization. One key feature, we utilize is that HPX resolves the data dependency and generates an asynchronous execution graph, which allows us to implicitly overlap the communication and the computation.
%HPX is in strict adherence to the C++ $11$ and C++ $17$ standard definitions.     

The paper is structured as follows: We present the related work in Section~\ref{s:RelatedWork}. In Section~\ref{s:nonlocalintroduction} we briefly introduce the nonlocal heat equation and in Section~\ref{s:Problemstatement} we highlight the key challenges in the development of \textit{distributed solver}. We discuss the key features of HPX used in our implementation in Section~\ref{s:hpxbasics} and then present our core technical contributions in Section~\ref{sec:Imp}. We propose a novel load balancing algorithm in Section~\ref{sec:loadBalancing}. We demonstrate the scaling results of our \textit{distributed solver} and load balancing algorithm in Section~\ref{sec:benchmarks} and conclude in Section~\ref{s:conclusion}.

%%%%%%%%%%%%%%%%%%%%%%%%%%%%%%%%%%
\section{Related Work}\label{s:RelatedWork}
%%%%%%%%%%%%%%%%%%%%%%%%%%%%%%%%%%
\noindent \textbf{Asynchronous many-task systems (AMT):}
Many asynchronous many-task (AMT) systems have been developed recently. However, we focus on those with distributed capabilities since the current work focuses on the domain decomposition and load balancing for the distributed case. Here, the most notable are: Uintah~\cite{germain2000uintah}, Chapel~\cite{chamberlain2007parallel}, Charm++~\cite{kale1993charmpp}, Kokkos~\cite{edwards2014kokkos}, Legion~\cite{bauer2012legion}, and PaRSEC~\cite{bosilca2013parsec}. According to \cite{thoman2018taxonomy}, only HPX has a C++ standard compatible API and has the highest technology readiness level.

\noindent \textbf{Load balancing:}
To the best of our knowledge, there are no load balancing algorithms specifically designed to handle the constraints of minimizing the data dependencies across multiple computational nodes in \textit{distributed solvers} for nonlocal models.~\cite{tree_match} proposed a load balancing algorithm in the Charm++ library that shares the underlying implementation concepts with HPX.~\cite{tree_match} proposed a hardware-topology aware load balancing to minimize the application communication costs. In a slightly different context of single computational node multi-core systems,~\cite{work_stealing} proposes to use the ideas of work-stealing and dynamic coarsening to improve the data locality on multiple cores while load balancing.

%%%%%%%%%%%%%%%%%%%%%%%%%%%%%%%%%%
\section{Brief Introduction to the Nonlocal Heat (Diffusion) Equation}
\label{s:nonlocalintroduction}
\begin{figure}[tb]
    \centering
    \includegraphics[scale=0.22]{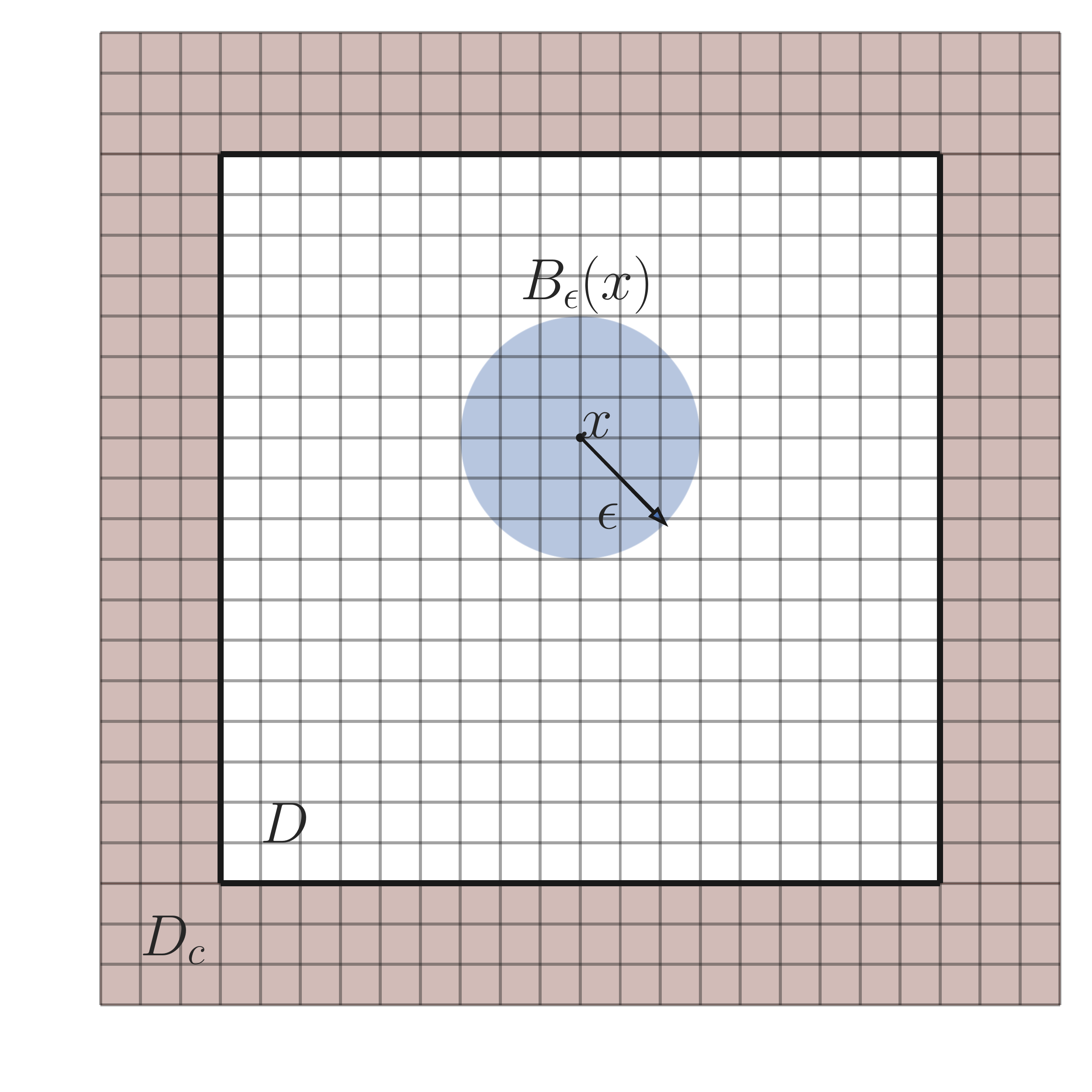}\vspace{-10pt}
    \caption{Material domain $D$ with the nonlocal boundary $D_c$ surrounding $D$. Figure shows typical material point $x\in D$ and its neighborhood region, a ball of radius $\epsilon$ centered at $x$, $B_\epsilon(x)$. We also show the spatial discretization of domains $D$ and $D_c$ through a uniform grid.}
    \label{fig:MaterialDomain}
\end{figure}

%%%%%%%%%%%%%%%%%%%%%%%%%%%%%%%%%%
In this section, we briefly introduce the nonlocal heat equation. We chose this model due to its simplicity, however, the algorithms and ideas discussed in this work should work for a more complex nonlocal fracture model such as peridynamics. We consider a two dimensional nonlocal diffusion equation for the temperature $u:[0,T]\times D \mapsto \R$ over an square domain $D = [0,1]^2$ for the time interval $[0,T]$. We impose a zero temperature boundary condition on the boundary of $D$ and specify a heat source $b:[0,T]\times D\mapsto \R$ as a function of points on the domain and time. Let $\epsilon > 0$ be the nonlocal length scale. The temperature field satisfies
\begin{align}\label{eq:DiffEqn}
\frac{\partial u(t,x)}{\partial t} &=  b(t,x) \notag \\
&\,+ c \int_{B_\epsilon(x)} J\left(\frac{|y-x|}{\epsilon}\right) (u(t,y) - u(t,x)) \msd y,
\end{align}
where $|y-x|$ denotes the Euclidean norm of vector $y-x$ in 2d, $B_\epsilon(x) = \{y: |y-x| \leq \epsilon\}$ the ball of radius $\epsilon$ centered at $x$, and $J = J(r)$ the influence function. We consider $J = 1$ for simplicity. Here, the constant $c$ is related to the heat conductivity $k$ of the material as follows
\begin{align}\label{eq:constc}
c &= \begin{cases}
\frac{k}{\epsilon^3 M_2}, \qquad \text{when dimension }d=1 \\
\frac{2k}{\pi\epsilon^4 M_3}, \qquad \text{when dimension }d=2,
\end{cases}
\end{align}
where $M_i = \int_D J(r) r^i \msd r$ is the $i^{\text{th}}$ moment of the influence function. \eqref{eq:constc} can be derived by substituting the Taylor series expansion, $u(t,y) = u(t,x) + \nabla u(t,x) \cdot (y-x) + \frac{1}{2}\nabla^2 u(t,x) \boldsymbol{\colon}(y-x)\otimes (y-x) + O(\epsilon^3)$, in \eqref{eq:DiffEqn} and comparing the Laplacian term with that from the classical heat equation. We refer to~\cite{burch2011classical} for more discussion on the nonlocal diffusion equation. The initial condition is given by
\begin{align}\label{eq:ic}
u(0,x) = u_0(x) \qquad \forall x\in D.
\end{align}
The boundary condition is typically prescribed over the region of finite area (in 2d) or volume (in 3d). Let $D_c = (-\epsilon, 1+\epsilon)^2 - D$ is the annuls square obtained by removing $D$ from the bigger square $(-\epsilon, 1+\epsilon)^2$, see Figure~\ref{fig:MaterialDomain}. We apply the zero temperature boundary condition on $D_c$, \emph{i.e.},
\begin{align} \label{eq:bc}
u(t,x) = 0 \qquad \forall x\in D_c \text{ and } \forall t\in [0,T].
\end{align}

To solve for the temperature $u$ for a given source $b$, we consider a finite difference in space and forward-Euler in time discretization of \eqref{eq:DiffEqn}. We discuss the discretization next.

%%%%%%%%%%%%%%%%%%%%%%%%%%%%%%%%%%
\subsection{Finite Difference Approximation}\label{ss:fd}
%%%%%%%%%%%%%%%%%%%%%%%%%%%%%%%%%%
We discretize the domain using the uniform grid with grid size $h>0$, see Figure~\ref{fig:MaterialDomain}. Let $D_h$ and $D_{c_h}$ denotes the spatial coordinates of grid points after discretization of $D$ and $D_c$. We consider an index set $K \subset \Z^2$ such that for $i\in K$, $x_i = h i \in D_h$. Similarly, we consider an index set $K_c \subset \Z^2$ such that for $i\in K_c$, $x_i = h i \in D_{c_h}$. Let $\{0, t_1 = \Delta t, t_2 = 2\Delta t, ..., t_N = N\Delta t\}$, such that $t_N \leq T$, is the discretization of the time interval $[0,T]$. Here $\Delta t$ is the size of the timestep.

Let $\hat{u}^k_i$ denote the temperature at grid point $i\in K\cup K_c$ and at time $t_k = k \Delta t$. Using the finite-difference approximation and forward-Euler time-stepping scheme, we write the discrete system of equations for $\hat{u}^k_i$, for all $i\in K$ and $1\leq k \leq N$,
\begin{align}\label{eq:ForwardFiniteDiff}
\dfrac{\hat{u}^{k+1}_i - \hat{u}^k_i}{\Delta t} &= b(t^k, x_i) \notag \\
&+ c \sum_{\substack{j \in K\cup K_c,\\
|x_j-x_i| \leq \epsilon}} J(|x_j - x_i|/\epsilon) (\hat{u}^k_j - \hat{u}^k_i) V_j.
\end{align}
The boundary condition over $D_c$ translates to
\begin{align*}
\hat{u}^k_i = 0, \qquad \forall k \in K_c, \, 0\leq k \leq N.
\end{align*}
To apply the initial condition, we set $\hat{u}^0_i = u_0(x_i)$ for all $i\in K$. In the above, $V_j$ is the area occupied by the grid point $j$ in the mesh. For the uniform discretization with the grid size $h$, we have $V_j = h^2$. 

\subsection{Exact Solution and Numerical Error}
\label{ss:exactsolution}

To test the implementation of the discretization scheme in Subsection~\ref{ss:fd}, we consider a method of constructing the exact solution. Let $$w(t,x) = \cos(2\pi t) \sin(2\pi x_1)  \sin(2\pi x_2)$$ when $x\in D$ and $w(t,x) = 0$ when $x\notin D$. We consider an external heat source of the form: 
\begin{align}\label{eq:exactSourceB}
b(t,x) &= \frac{\partial w(t,x)}{\partial t} \notag \\
&\, -  c \int_{B_\epsilon(x)} J(|y-x|/\epsilon) (w(t,y) - w(t,x)) \msd y .
\end{align}
With $b$ given by above and the initial condition $$u_0(x) = w(0,x) = \sin(2\pi x_1)\sin(2\pi x_2),$$ we can show that $u(t, x) = w(t,x)$ is the exact solution of \eqref{eq:DiffEqn}. Next, we define the numerical error. 

Suppose $\bar{u}(t,x)$ is the exact solution and $\hat{u}^k_i$ for $0\leq k \leq N$ and $i\in K$ is the numerical solution. The  error at time $t^k$ is taken as
\begin{align}\label{eq:err}
e^k &= h^d \sum_{i\in K} |\bar{u}(t^k, x_i) - \hat{u}^k_i|^2,
\end{align}
where $d=1,2$ is the dimension. The total error can be defined as the sum of errors $e^k$, \emph{i.e.}, $e= \sum_{0\leq k\leq N} e^k$. 

%%%%%%%%%%%%%%%%%%%%%%%%%%%%%%%%%%
\section{Problem Statement and Formalism}\label{s:Problemstatement}
%%%%%%%%%%%%%%%%%%%%%%%%%%%%%%%%%%

Fully parallelizing a serial implementation of the nonlocal equation to a distributed version deployed on an Asynchronous Many-Task System (AMT) involves numerous challenges. These challenges are primarily related to the work distribution across multiple computational nodes, minimizing the idle time spent on data exchange among computational nodes and ensuring maximum efficiency across all computational nodes. Designing and implementing a \textit{distributed solver} that addresses and overcomes the challenges listed below is very critical to ensure optimal performance:-
\begin{enumerate}[leftmargin=15pt]
    \item \textit{\textbf{Mesh partitioning}} -- breaking down the main problem into smaller sub-problems that can be distributed across multiple computational nodes. Each computational node may contain multiple such sub-problems. This simple idea of unitized work greatly helps to simplify the work distribution and load balancing in AMTs, where there are multiple computational nodes within a cluster and each computational node consists of multiple CPUs. In such a scenario, each of the sub-problems can be easily assigned to different threads within a single computational node to utilize multiple CPUs on a single computational node.
    \item \textit{\textbf{Minimizing data exchange}} -- distributing the sub-problems among multiple computational nodes to achieve minimum data dependencies. The efficiency of a \textit{distributed solver} is limited by the data dependency across the computational nodes. Minimum data dependency across the computational nodes ensures minimum data exchange time across different computational nodes and better scaling. 
    \item \textit{\textbf{Hiding data exchange time}} -- doing useful computation while waiting for the data from the neighboring computational nodes. Despite the optimal work distribution, the computational nodes need to exchange data required for the computation of the nonlocal solver. To avoid keeping the computational nodes idle while the data is being exchanged, it is possible to perform computations on portion of owned domain that do not depend on the data of other nodes. This asynchronous-style computation helps us to hide the data exchange time and to ensure optimal performance of the \textit{distributed solver}.
    \item \textit{\textbf{Load balancing}} -- redistribution of sub-problems across multiple computational nodes according to their load to minimize the waiting time across the faster nodes. Compute capacity of the individual computational nodes may vary with time, either due to scheduling of some other task or due to the intrinsic behaviour of the nonlocal model. In such a scenario, the updated compute capacity of individual nodes needs to be accounted and the load needs to be balanced so that the faster nodes do not sit idle.
\end{enumerate}

We describe our solution to address all challenges above in Section~\ref{sec:Imp} and propose a novel load balancing algorithm in Section~\ref{sec:loadBalancing}. \\

\begin{figure}[tb]
\begin{center}
    \includegraphics[scale=0.2]{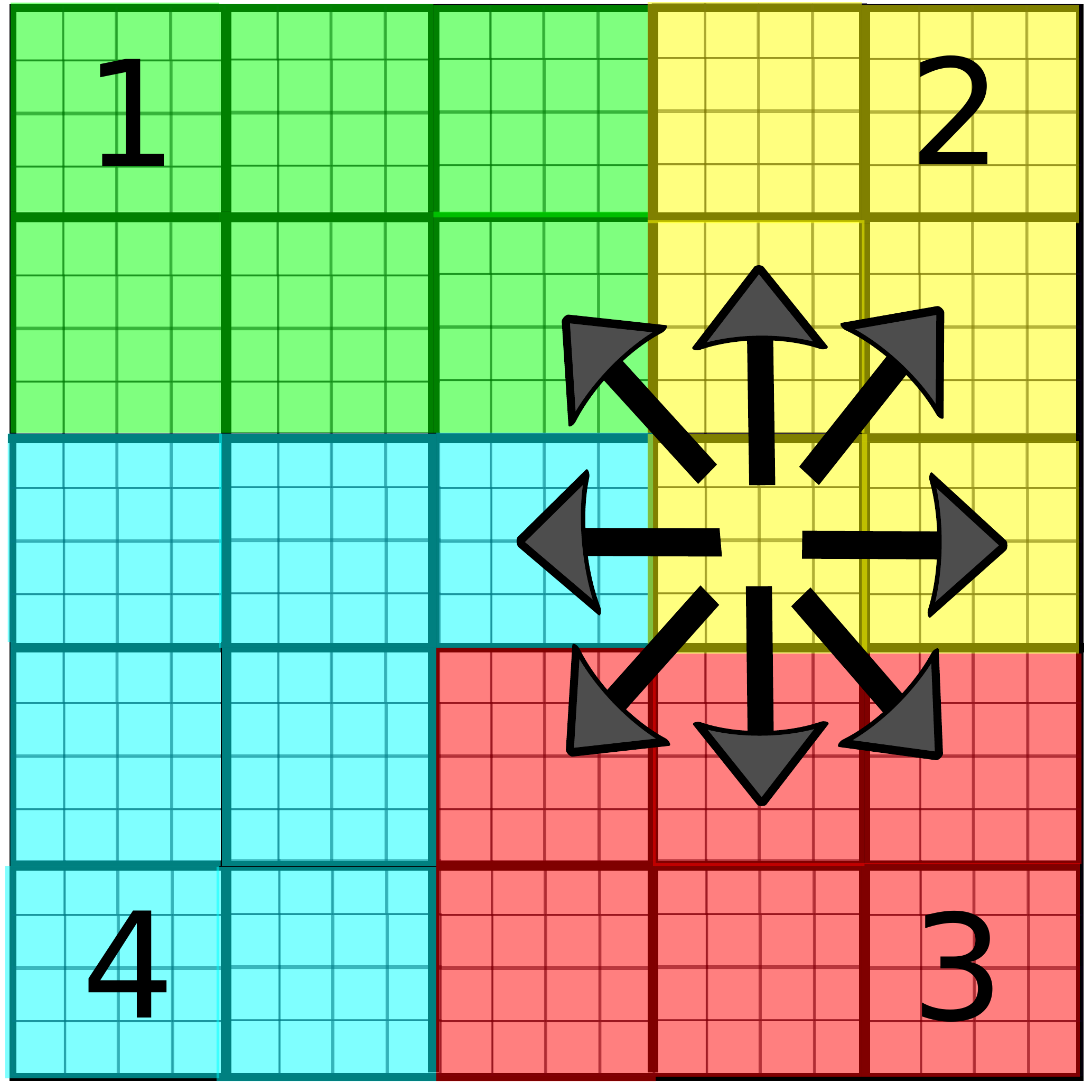}
    \caption{Discretization of the domain (grid with smaller spacing) and $\SD$s (grid with larger spacing, note the thicker lines). Here, we consider 4 compute nodes and 25 square $\SD$s. Color indicates the owner of the $\SD$s. As we can see a typical $\SD$ consists of $4\times 4$ $\DP$s (discretized points). To compute the updated temperature at $\DP$s within an $\SD_i$, the data from neighboring $\SD$s within the $\epsilon$ distance is required. If the size of the $\SD$ is bigger than $\epsilon$, we see that any $\SD$ needs to communicate only with the immediate neighboring $\SD$s.}
\label{fig:DataExchange}
\end{center}
\end{figure}

%\noindent \textbf{\textit{FORMALISM:}}
\noindent \textbf{{Formalism}}

We define the following terms used throughout this and the following sections:
\begin{itemize}[leftmargin=15pt]
    \item \textit{\textbf{Distributed solver}} -- the time-stepping scheme \eqref{eq:ForwardFiniteDiff} where we start with the given initial temperature at discrete points, $\hat{u}^0$, in the domain and apply \eqref{eq:ForwardFiniteDiff} to compute the temperature at successive times $t_1 = \Delta t, t_2 = 2\Delta t, .., t_k = k\Delta t$. At each timestep $k$, the right-hand side term of \eqref{eq:ForwardFiniteDiff} must be computed first to compute the temperature at the next timestep $t_{k+1}$. This involves an underlying challenge of the data dependency for all the points within $\epsilon$ distance of a given point $x_i$. Since $\epsilon$ is typically higher than $2h$ where $h$ is the grid size, each computational node needs to scatter and gather the temperature values at the degree of freedoms owned by the neighboring computational nodes.
    \item \textit{\textbf{Sub-problem}} ($\SP$) -- the region of the material domain managed by a computational node; \emph{i.e.} the computational node is the owner of the degrees of freedom in that region and computes the temperature following scheme \eqref{eq:ForwardFiniteDiff}. For example, in Figure~\ref{fig:DataExchange}, the region colored with green, yellow, cyan, and red shows four $\SP$s.
    \item \textit{\textbf{Sub-domain}} ($\SD$) -- the region of the material domain that is processed and computed independently within a given computational node. $\SD$s are subsets of $\SP$ in a given node; \emph{i.e.} $\SP$ is further divided into multiple $\SD$s. In this work, we always consider a square sub-domain. By the size of $\SP$, we mean the number of $\SD$s it consists of. In Figure~\ref{fig:DataExchange}, the $\SD$s are grids with larger spacing (see thick lines); for instance, node $1$ has 6 $\SD$s.
    \item \textit{\textbf{Discretized point}} ($\DP$) -- each $\SD$ is responsible for the discrete points within it. To be able to apply \eqref{eq:ForwardFiniteDiff} and compute the temperature at the discrete points in a given $\SD_i$, $\SD_i$ will have to exchange the data from neighboring $\SD$s. When all neighboring $\SD$s of $\SD_i$ are in the same computational node as $\SD_i$, no special data exchange method is required. However, when some $\SD$s are in neighboring computational nodes, a proper communication method is required. For instance, in Figure~\ref{fig:DataExchange}, the green $\SD$ (owned by node $1$), on top and near yellow region, depends on the data owned by node $2$.
\end{itemize}

%%%%%%%%%%%%%%%%%%%%%%%%%%%%%%%%%%
\section{HPX Basics}
\label{s:hpxbasics}
%%%%%%%%%%%%%%%%%%%%%%%%%%%%%%%%%%
\begin{figure}[tb]
    \centering
  \includegraphics[width=0.75\linewidth]{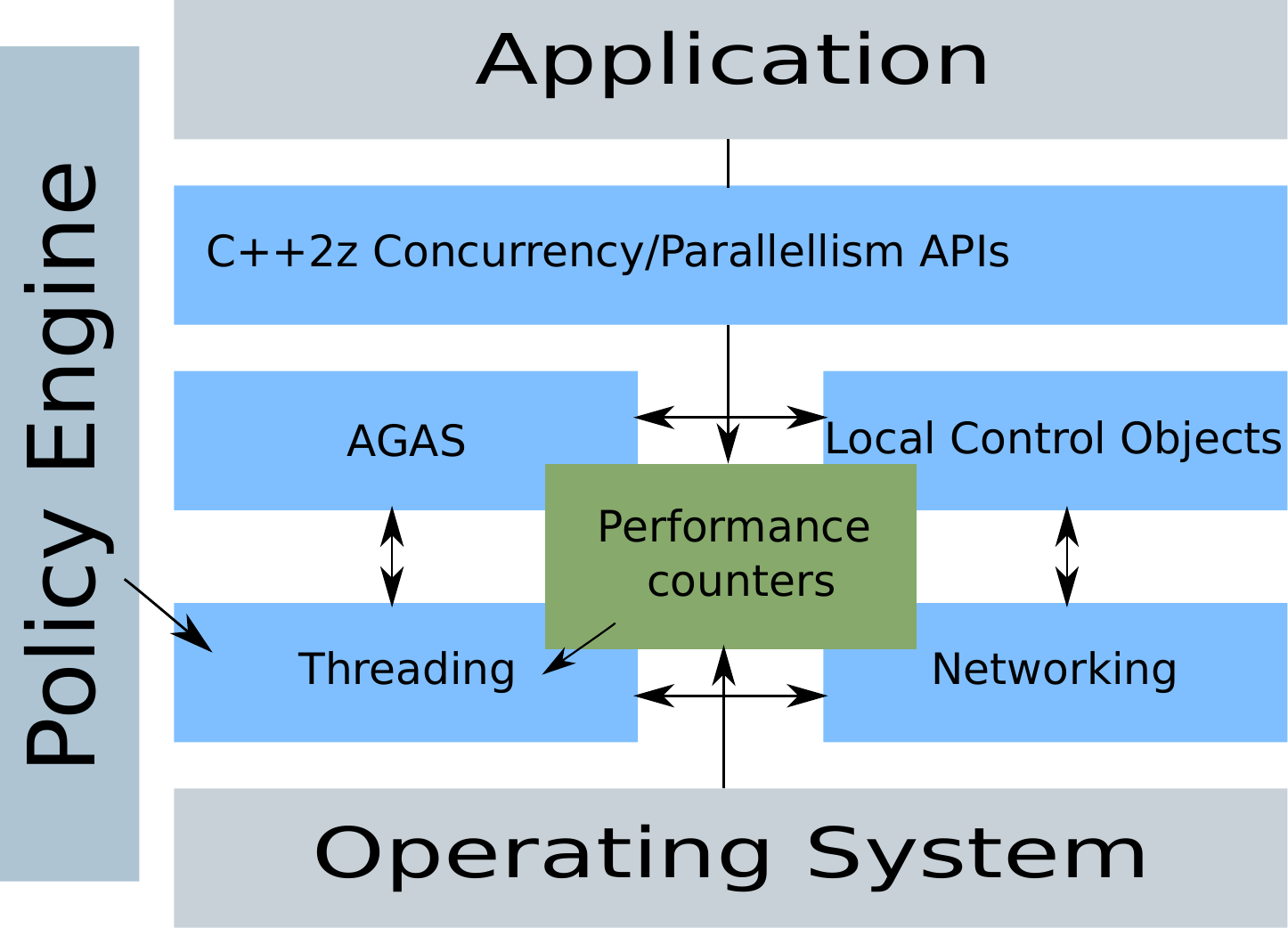}
    \caption{Sketch of the HPX's architecture and its components: Local Control Objects (LCOs), Threading sub-system, Networking, performance counters, and Active Global Address Space (AGAS). We briefly introduce the performance counter and the local control objects components in this section. For all other components, we refer to~\cite{Kaiser2020}. Adapted from~\cite{Kaiser2020}.}
    \label{fig:hpx:components}
\end{figure}

Figure~\ref{fig:hpx:components} sketches HPX's architecture and its components. 
We briefly introduce the local control objects and the performance counters that we use in our implementation.
For all other components, we refer to~\cite{Kaiser2020}. The performance counters component provides a uniform API of a globally accessible counter to access the performance in-situ~\cite{grubel2016dynamic}. All counters are registered within the Active Global Address Space (AGAS)~\cite{amini2019assessing} to poll the counters during the run time of the application. 
%These counters are used in the presented load balancing algorithm. 

The Local Control Objects (LCOs) component provides the features \lstinline|hpx::async| and \lstinline|hpx::future| for the asynchronous execution and synchronization.  Listing~\ref{listing:futures} shows how to compute $a + b + c + d$ asynchronously using HPX. In Line~\ref{listing:futures:func} the function to compute two integers is defined. In Lines~\ref{listing:futures:async1}--\ref{listing:futures:async2} the function \lstinline|add| is launched asynchronously using \lstinline|hpx::async| which means the function is executed on one thread and immediately returns a \lstinline|hpx::future<int>| which contains the result of the computation once the thread is finished. Thus, the second function call is launched and the two function calls are executed concurrently. In Line~\ref{listing:futures:sync} a blocking synchronization of the two asynchronous function calls is needed to obtain the results using the \lstinline{.get()} function.

\begin{lstlisting}[language=C++, caption=Example to illustrate the usage of \lstinline|hpx::future| and \lstinline|hpx::async|, label=listing:futures,escapechar=|, basicstyle=\fontsize{7}{3}\selectfont]
1: // Function to add two integers
2: int add (int one, int second){ |\label{listing:futures:func}|
3:     return one + second;
4: }
5: // Launch two function calls asynchronously 
6: hpx::future<int> a_add_b = hpx::async(add, a, b);|\label{listing:futures:async1}|
7: hpx::future<int> c_add_d = hpx::async(add, c, d);|\label{listing:futures:async2}|
8: // Synchronization to compute the result 
9: int result = a_add_b.get() + c_add_d.get(); |\label{listing:futures:sync}|
\end{lstlisting}

%%%%%%%%%%%%%%%%%%%%%%%%%%%%%%%%%%
\section{Distributed Solver Implementation}
\label{sec:Imp}
%%%%%%%%%%%%%%%%%%%%%%%%%%%%%%%%%%
In this section, we present our approach to solve the various challenges associated with the \textit{distributed solver} discussed in Section~\ref{s:Problemstatement}.
To understand the challenges and compute the gain in going from a serial (single computational node and single-threaded) implementation to a fully distributed and load-balanced implementation, we first implemented a single-threaded version. Second, we extended the serial implementation to a multi-threaded version using asynchronous execution, \emph{e.g.,}\ futurization. 
Third, we implemented a fully distributed asynchronous solver for \eqref{eq:ForwardFiniteDiff}. In this extension, we rely on the HPX's implementation of the distributed asynchronous computation. 
We demonstrate the schematics of the fully distributed and load-balanced time-stepping scheme of nonlocal models~\eqref{eq:ForwardFiniteDiff} in Figure~\ref{fig:Summary}. In Figure~\ref{fig:Summary}(right), various components that tackle the challenges listed in Section~\ref{s:Problemstatement} and their interactions are shown. We now discuss our approach that address the challenges described in Section~\ref{s:Problemstatement}:

\begin{figure}[tb]
\begin{center}
    \includegraphics[scale=0.4]{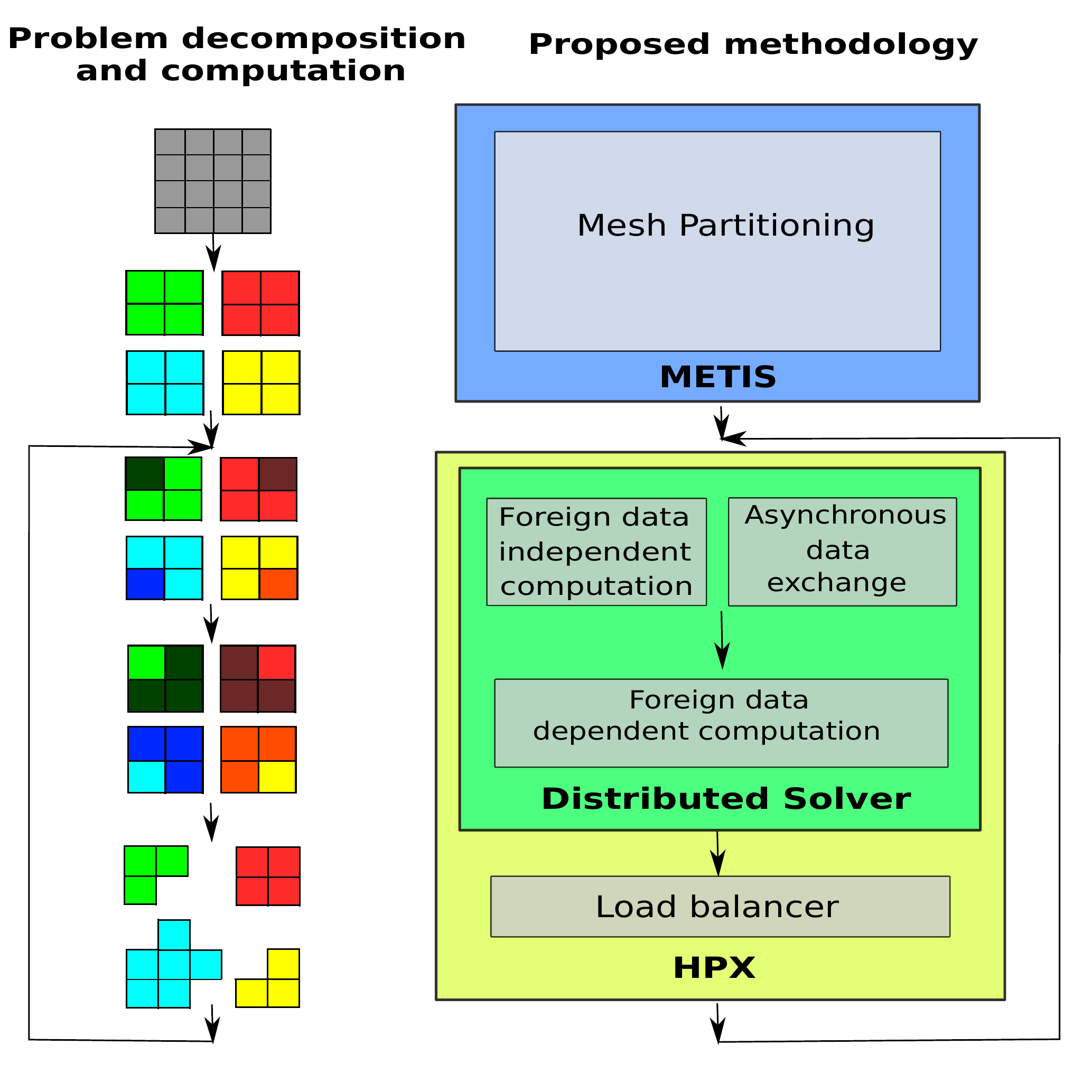}
    \caption{Sketch of the proposed distributed framework (on right) and the decomposition and computation problem flow in a distributed cluster (on left). {\bf Right}: For the decomposition of the domain and the distribution of the problems across computational nodes, METIS library was utilized. Foreign data (foreign data is the data that is not available on the current node) independent computation is done while waiting for the data to be available from the neighboring nodes. Foreign data-dependent computation is done once the data from the neighboring nodes is available. Our load balancing algorithm ensures that all the computational nodes have the $\SP$ (sub-problem) size in proportion to their computational power. {\bf Left}: The problem is first decomposed into multiple $\SP$s distributed to nodes (colors show the node responsible for the $\SP$). In the second step, see the third figure from the top, while waiting for data from the neighboring nodes, we perform computation on those $\SD$s (sub-domains) which do not depend on the data of neighboring processors; the dark-colored square indicates that computation is being performed on the $\DP$s (discretized points) within it. In the third step, see the fourth figure from the top, we now perform computation on those regions of $\SP$ that depended on the neighboring nodes' data; the dark-colored squares are now different. Finally, at the end of the timestep, we check for the load on each compute node, and if needed redistribute the $\SD$s (\emph{i.e.} change the $\SP$ of individual nodes) by applying the load balancing step discussed in Section~\ref{sec:loadBalancing}.}
\label{fig:Summary}
\end{center}
\end{figure}

% %%%%%%%%%%%%%%%%%%%%%%%%%%%%%%%%%%
% \subsection{Shared memory implementation}\label{ss:SharedMemoryImp}
% %%%%%%%%%%%%%%%%%%%%%%%%%%%%%%%%%%
% In this section we demonstrate the speedup based on instruction level parallelism. The main idea is to distribute the computation of the entire domain among multiple threads located on the same node. The data exchange among the threads is using the shared memory. Common data structures are shared among the threads on  a single node. We propose in the following section to partition the mesh into coarser units. Each thread will be responsible for computing the temperature for the part of the domain owned by itself. Please refer to the next section on Mesh partitioning which describes the partitioning of the mesh into coarser units for efficient parallel execution.

% %%%%%%%%%%%%%%%%%%%%%%%%%%%%%%%%%%
\subsection{Mesh Partitioning}
\label{ss:meshPartitioning}
% %%%%%%%%%%%%%%%%%%%%%%%%%%%%%%%%%%
We propose a simple decomposition of the complete square domain into smaller squares ($\SD$s). As shown in Figure~\ref{fig:DataExchange}, we divide the discretized square domain into 5 x 5 $\SD$s, \emph{i.e.}, total 25 $\SD$s are created. Each $\SD$ consists of 4 x 4 $\DP$s and is responsible for the computation associated to these $\DP$s. We consider the computation of an $\SD$ as a unit of work; the size of $\SD$ controls the communication burden and the size of work (larger $\SD$ will have more $\DP$s and hence more computation) and therefore the size of an $\SD$ can be tuned to achieve maximum performance.

\subsection{Minimizing the Data Exchange}\label{ss:DomainDecompMetis}
%%%%%%%%%%%%%%%%%%%%%%%%%%%%%%%%%%

The efficiency of a \textit{distributed solver} is limited by the data dependency across the computational nodes. For instance, in Figure~\ref{fig:DataExchange}, the $\SD$s belonging to a computational node \textit{2} depend on the data with the neighboring $\SD$s belonging to computational nodes \textit{1}, \textit{3} and \textit{4}. We propose to use the METIS library~\cite{metis_lib} to address this challenge. METIS is a set of serial programs for providing fast and high-quality partitioning of finite element meshes and graphs. Specifically, we use the \lstinline|METIS_PartMeshDual| function which ensures that the resulting partition is optimal and results in minimum data exchange across the computational nodes during the execution. For instance, we use \lstinline|METIS_PartMeshDual| to distribute 25 $\SD$s across 4 nodes as shown in Figure~\ref{fig:DataExchange}. The contiguous partitioning ensures that most of the bordering elements would exchange data with the $\SD$s belonging to the same computational node, thus, reducing the data exchange.

% %%%%%%%%%%%%%%%%%%%%%%%%%%%%%%%%%%
\subsection{Hiding the Data Exchange Time}
\label{ss:dataExchange}
% %%%%%%%%%%%%%%%%%%%%%%%%%%%%%%%%%%

\begin{figure}[tb]
\begin{center}
    \includegraphics[scale=0.23]{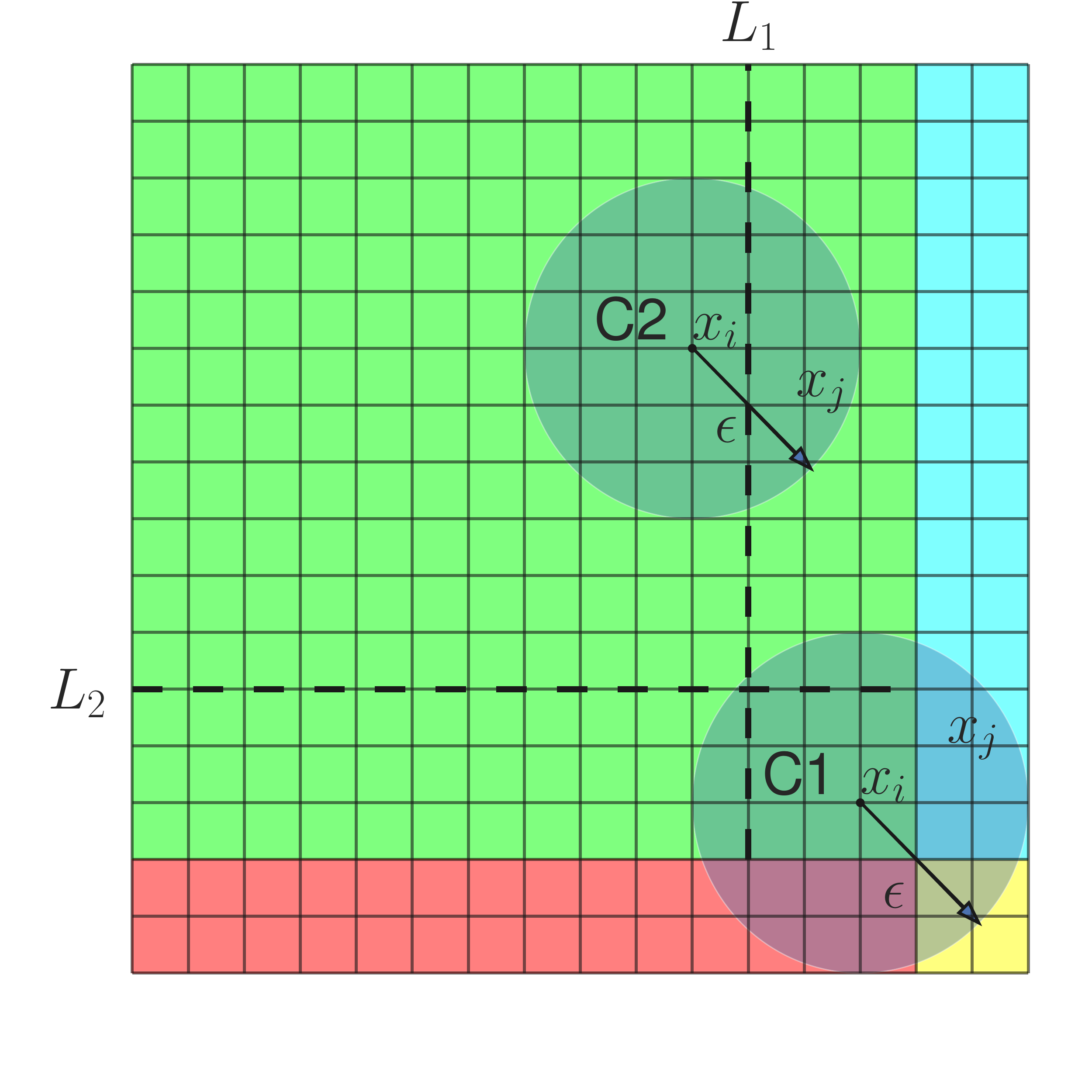}\vspace{-20pt}
    \caption{Different cases of computation in one $\SD$ (green square). The neighboring $\SD$s owned by different nodes is also shown. The $\DP$s in green $\SD$ can be divided into two parts: 1) $\DP$s on right of line $L_1$ or bottom of $L_2$ (see typical point $x_i$ near label {\bf C1}) and 2) remaining $\DP$s in green $\SD$ (see typical point $x_i$ near label {\bf C2}). 
    At timestep $k$, we can perform computation on $\DP$s independently for case 2. Whereas for $\DP$s associate to case 1 the computation depends on the data from neighboring nodes.}
    \label{fig:Case2}
\end{center}
\end{figure}

To hide the data exchange time, we propose to divide the set of $\DP$s of any $\SD$ into two cases (also see Figure~\ref{fig:Case2}):

\begin{itemize}[leftmargin=15pt]
    \item \textit{\textbf{Case 1}} consists of $\DP$s that depend on the data from $\SD$s of other nodes. Consider a point $x_i$ near label {\bf C1} in Figure~\ref{fig:Case2}; the $\epsilon$ neighborhood of $x_i$ consists of some $\DP$s on $\SD$s of other computational nodes. At every timestep, to compute the right-hand side in \eqref{eq:ForwardFiniteDiff}, the updated temperature at of $\DP$s such as $x_j$ of neighboring $\SD$s must be collected. 

\item \textit{\textbf{Case 2}} consists of the remaining $\DP$s. The computation associated with these $\DP$s is independent of the $\DP$s belonging to other computational nodes. 

\end{itemize}

At every timestep, we propose to compute the data for the $\DP$s belonging to \textit{Case 2} first, while the data for $\DP$s belonging to \textit{Case 1} becomes available. This ordering ensures an effective hiding of the data sharing time for the points corresponding to  \textit{Case 1}.

%% Add a partition figure here

%%%%%%%%%%%%%%%%%%%%%%%%%%%%%%%%%%
\section{Load Balancing}
\label{sec:loadBalancing}
%%%%%%%%%%%%%%%%%%%%%%%%%%%%%%%%%%
The issue of load balancing often becomes crucial in nonlocal models, especially in nonlocal fracture models~\cite{diehl2019review}. In fracture/crack problems, the computation in regions containing the crack is different from the region not containing it; the crack line (in 2d or surface in 3d) divides the two regions such that the points on either side of the crack line do not interact with each other. In our terminology, the $\SD$s which contain the portion of crack will have reduced computational burden as compared to $\SD$s not containing the crack. Another instance where load-balancing could be crucial is when the computational capacity of a node is time-dependent.
%A similar type of load imbalance is observed when the computational power of a node is time-dependent, which calls for an efficient load balancing algorithm. 
To the best of our knowledge, there are no well-known load balancing algorithms for nonlocal models in AMT systems. In this section, we propose a novel load balancing algorithm useful for the cases discussed above. The only assumption we make is that the data exchange times are negligible compared to the amount of time spent on the computation. Our method of hiding the data exchange times in  Subsection~\ref{ss:dataExchange} makes this assumption realistic and reasonable. We now discuss the key aspects of the load balancing algorithm:

%While in this work we restricted to nonlocal heat equation, our goal was to design a parallel framework that can also apply to other class of nonlocal models such as peridynamics, nonlocal corrosion model, granular media models based on peridynamics, etc. 

\paragraph{\textbf{Calculating the Load Imbalance}}

To measure the load imbalance, we use \lstinline|hpx::performance_counters::busy_time| which reports the fraction of the time the node was busy doing computation against the total time the node was active. Between successive load balancing iterations, the performance counter is reset to have the same total time span for all the computational nodes in the cluster. Significantly different busy times for the different computational nodes indicate a load imbalance. In an ideal case, the busy time should be the same for all nodes. To achieve the close to perfect load-balanced state, it is necessary to assign $\SD$s to individual nodes based on their compute capacity. To measure the compute capacity of a particular computational node $N_i$, we consider the formula:
\begin{align}\label{eq:PowerBusyTime}
\mathrm{Power}(N_i) =\frac{\bar{\SD}(N_i)}{\text{Busy Time($N_i$)}} ,
\end{align}
where $\bar{\SD}(N_i)$ denotes the number of $\SD$s on node $N_i$. Clearly, a more powerful node can handle a larger number of $\SD$s. Thus, $\mathrm{\textit{Power}}$ can be seen as an accurate measure to quantify the compute capacity. 
We calculate the load imbalance in terms of $\SD$s using the following formula:
\begin{align}\label{eq:LoadImbalance}
\mathrm{LoadImbalance}(N_i) = E(N_i) - \bar{\SD}(N_i) ,
\end{align}
where $E(N_i)$ is given by:
\begin{align}\label{eq:ExpectedLoad}
E(N_i) = \text{Total no. of $\SD$} \times \space \frac{\mathrm{Power}(N_i)}{\sum_j \mathrm{Power}(N_j)} .
\end{align}

\begin{figure}[tb]
\begin{center}
    \includegraphics[scale=0.28]{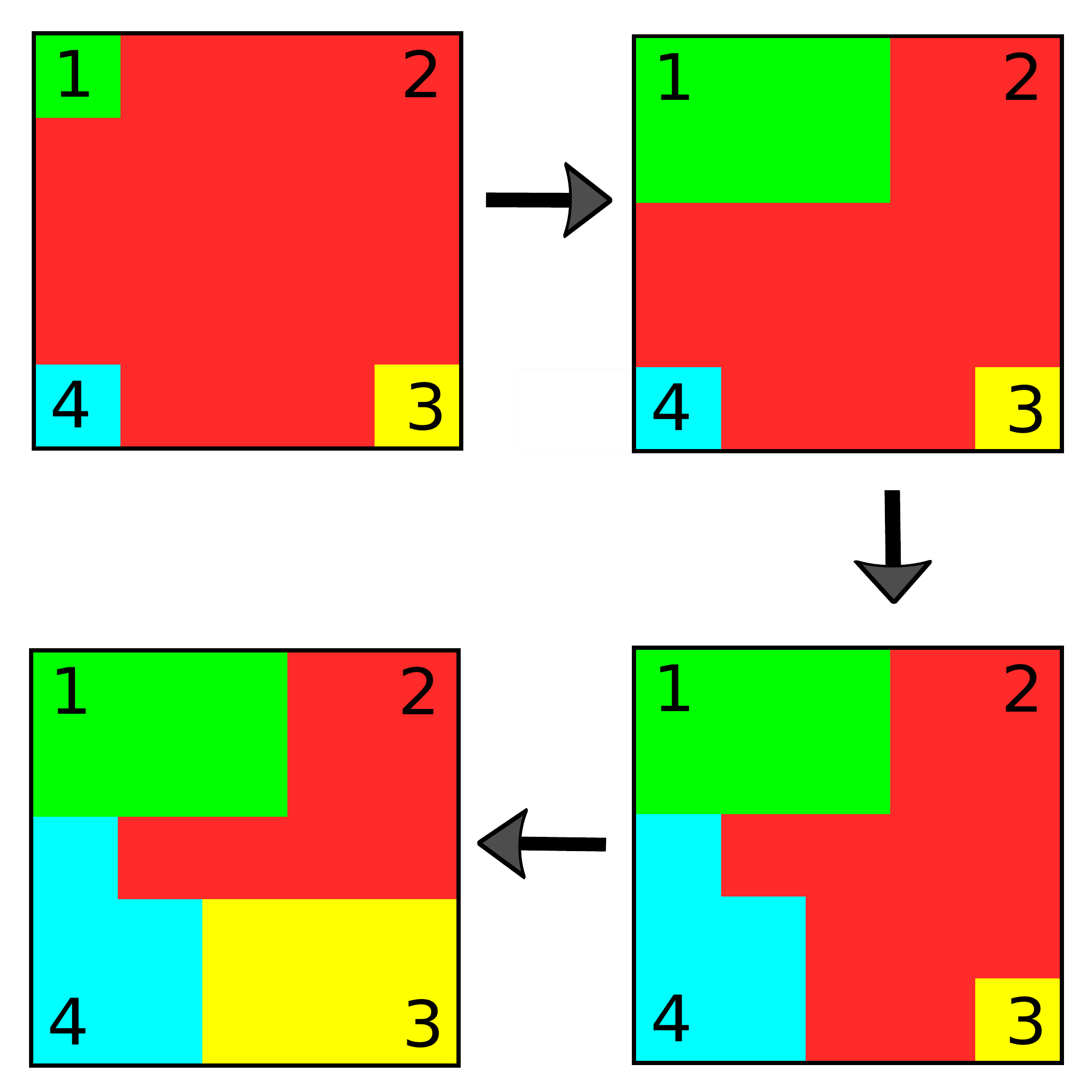}
    \caption{Redistribution of the \textit{sub-domain}s ($\SD$s) among the computational nodes to balance the load in the topological order. Node $1$ borrows the data from the non-visited node $2$, followed by nodes $4$ and $3$. Each node borrows $\SD$s uniformly in all the directions to retain a contiguous locality of the $\SD$s.}
\label{fig:LoadBalanceSteps}
\end{center}
\end{figure}

\begin{figure}[tb]
\begin{center}
    \includegraphics[scale=0.28]{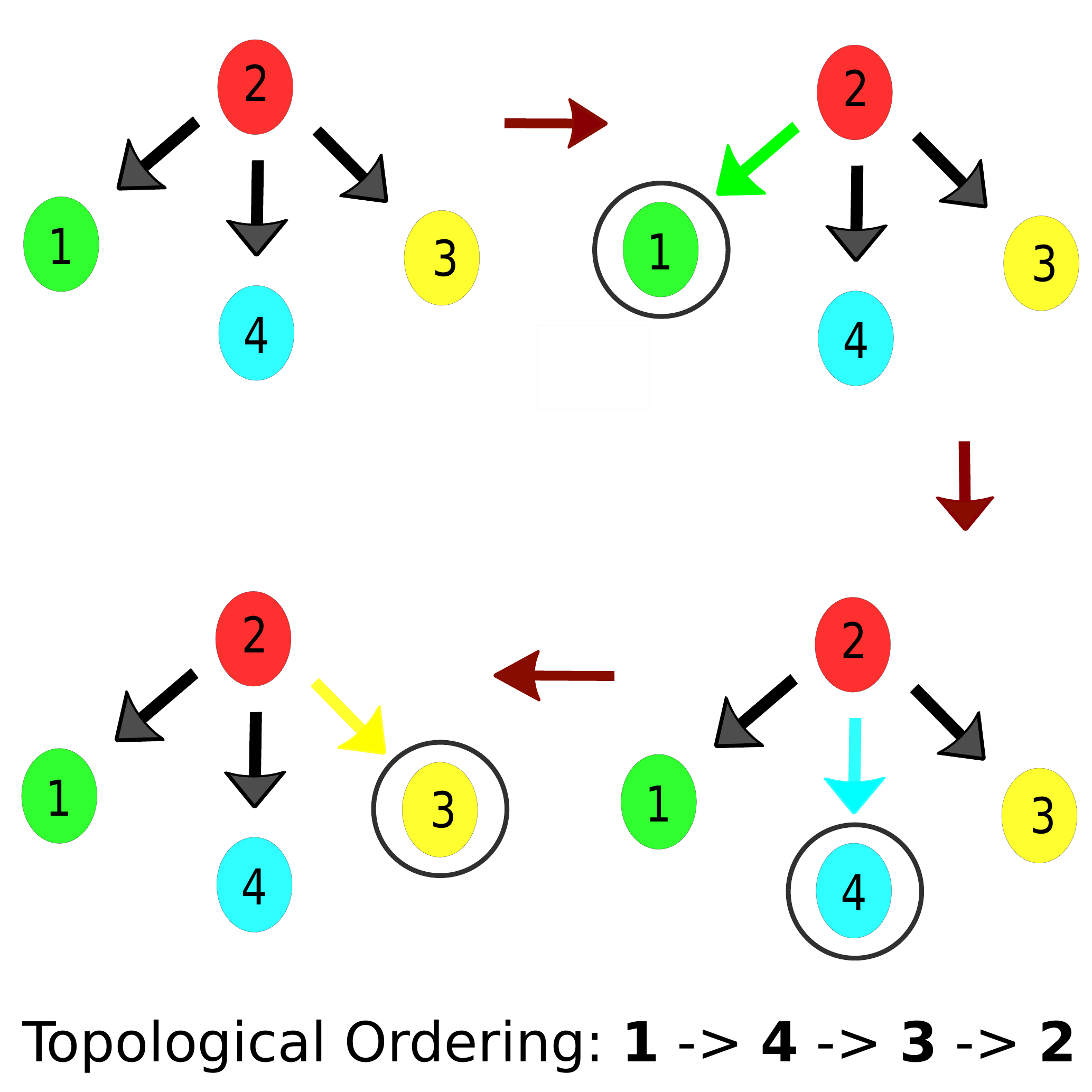}
    \caption{Data exchange among different pair of the nodes of the data dependency tree in the topological order. Topological ordering generated for the above figure is 1 $\rightarrow$ 4 $\rightarrow$ 3 $\rightarrow$ 2. Hence, node 1 borrows $\SD$s from its neighbouring node 2 to balance the load. Similarly, node 4 and then node 3 borrow $\SD$s from their neighbouring node.}
\label{fig:GraphEvolution}
\end{center}
\end{figure}

\paragraph{\textbf{Balancing the Load}}
We want to distribute the total work, \emph{i.e.}, all $\SD$s, keeping in mind the compute capacity, see \eqref{eq:LoadImbalance}, of the individual nodes.
% The total work, \emph{i.e.}, all the $\SD$s present in the domain need to be distributed among all computational nodes in the same ratio as their computing ability as indicated by \eqref{eq:LoadImbalance}, to ensure load balancing in the distributed solver. 
Towards this, we propose our novel load balancing algorithm; 
% We propose a novel load balancing algorithm to redistribute the load among various nodes in case of load imbalance. 
the main idea is to borrow/lend $\SD$s from/to a computational node, that owns the adjacent $\SD$s. For instance, in Figure~\ref{fig:LoadBalanceSteps}(top-left), the computational node $1$ borrows $\SD$s from $2$ to reduce the compute burden from $2$. Note that the $\SD$s belonging to $1$ share their boundaries with the $\SD$s of $2$. This helps in preserving a contiguous locality of $\SD$s, thereby minimizing the data exchange across computational nodes. 

The load balancing algorithm is presented briefly in Algorithm~\ref{alg:LoadBalance}. We calculate the load imbalance for each computational node by~\eqref{eq:LoadImbalance} in Lines~\ref{line:BeginLoadImbalance}--\ref{line:EndLoadImbalance}. In Lines~\ref{line:StartDependencyTree}--\ref{line:EndDependencyTree}, we model the data dependencies across computational nodes using a tree $T$; each node $N_i$ in the tree has a one-to-one correspondence with the computation node, and, an edge $e$ between two nodes of the tree denotes the data dependencies among them. An edge between nodes $N_i$ and $N_j$ can exist only if there is an $\SD$ in one of the nodes such that $\SD$ has data dependencies with $\SD$s of the other node. For instance, the scenario in Figure~\ref{fig:LoadBalanceSteps}(top-left) translates to the tree shown in Figure~\ref{fig:GraphEvolution}(top-left). Note that there are multiple such trees possible; we use one of the possible trees. In Line~\ref{line:TopologicalOrdering} of Algorithm~\ref{alg:LoadBalance}, we obtain an ordering of the nodes to perform the data redistribution in a \textit{least data-dependency first} fashion. 
The topological ordering used for the tree in Figure~\ref{fig:GraphEvolution} is $1 \rightarrow 4 \rightarrow 3 \rightarrow 2$. Coming back to the algorithm, in Lines~\ref{line:StartDependencyTree} -~\ref{line:EndDependencyTree}, we perform the actual data redistribution among the nodes in the topological order. Topological ordering helps to borrow / lend $\SD$s from different computational nodes optimally without unbalancing the already balanced nodes. Figure~\ref{fig:GraphEvolution} shows the redistribution of the $\SD$s among the tree nodes in the topological order and Figure~\ref{fig:LoadBalanceSteps} shows the change in the $\SD$ distribution across nodes. Note that each node $N_i$  borrows / lends $\SD$s to all the non-visited adjacent nodes $N_j$, uniformly in all the spatial directions in the domain. METIS generates an optimal distribution of $\SD$s across computational nodes for minimum data exchange. It is important to preserve the shape of $\SP$s obtained using METIS mesh partitioning in Subsection~\ref{ss:DomainDecompMetis} to obtain the best scaling. For instance, the node $1$ in Figure~\ref{fig:LoadBalanceSteps} borrows $\SD$s of $2$ from all spatial directions to minimize the data dependencies. At the end of the load balancing iteration, we reset the performance counters to obtain the correct performance measurements for the new load distribution for the next load balancing step; see Line~\ref{line:ResetCounters}.

% Insert the algorithm
\begin{algorithm}[tb]
	\caption{Load Balancing Algorithm}
	\label{alg:LoadBalance}
	\begin{algorithmic}[1]
	    \STATE \codecomment{Let $N$ is the number of computational nodes}
	    \STATE \codecomment{Compute number of $\SD$s (sub-domains) on each node}
	    \label{line:BeginLoadImbalance}
		\STATE compute NumSubDomains[i] for $i\in [0,N-1]$
		\STATE \codecomment{Compute computational power of node using \eqref{eq:PowerBusyTime}}
		\STATE compute Power[i] for $i\in [0,N-1]$ 
		\STATE \codecomment{Compute expected number of $\SD$s using \eqref{eq:ExpectedLoad} and $\mathrm{Power}$}
		\STATE compute ExpSubdomains[i] for $i\in [0,N-1]$
		\STATE \codecomment{Compute load imbalance using \eqref{eq:LoadImbalance}. Positive value indicate the load on node is smaller and negative otherwise.}
		\FOR {each integer $i \in [0, N-1]$}
		    \STATE LoadImbalance[i] = ExpSubdomains[i]
		    \STATE \space \space \space \space \space - NumSubDomains[i]
		\ENDFOR
		\label{line:EndLoadImbalance}
		\STATE \codecomment{Set minimum imbalanced load node as root R of tree T}
		\label{line:StartDependencyTree}
		\STATE Root R = argmin(LoadImbalance)
		\STATE \codecomment{Each node is represented by some node in the tree T}
		\STATE \codecomment{An edge e between two nodes exists if there is $\SD$ that is in one of the node and is adjacent to the $\SP$ (sub-problem, set of $\SD$s) of other node}
		\STATE Tree T = construct$\_$tree(R)
		\STATE \codecomment{Give root R and tree T, get$\_$topological$\_$sort(R, T) returns node ids to be processed in next step}
		\label{line:EndDependencyTree}
		\STATE orderedNodes[N] = get$\_$topological$\_$sort(R, T)
		\label{line:TopologicalOrdering}
		\STATE \codecomment{Each node n borrows $\SD$s only from non-visited nodes to increase its territory (set of $\SD$s) uniformly in all the directions}
		\FOR {each node $i \in$ orderedNodes$[0, N-1]$}
		\label{ine:StartLoadDistribution}
		    \IF{LoadImbalance[$i$] == 0}
		        \STATE continue        
		    \ENDIF
		    \STATE \codecomment{Compute list of non-visited adjacent nodes of node $i$}
		    \STATE \codecomment{suppose \text{AdjacentNonVisitedNodes} contains L nodes}
		    \STATE compute AdjacentNonVisitedNodes
		    \STATE \codecomment{Number of $\SD$s to borrow from each adjacent node}
		    \STATE XchngNum = LoadImbalance[$i$] / L
		    \FOR {each node $m \in$ AdjacentNonVisitedNodes($i$)}
		        \STATE LoadImbalance[$m$] -= XchngNum
		    \ENDFOR
		    \STATE LoadImbalance[$i$] = 0
		\ENDFOR
		\label{ine:EndLoadDistribution}
		\STATE reset$\_$all(\lstinline|hpx::performance_counters::busy_time|)
		\label{line:ResetCounters}
	\end{algorithmic}
\end{algorithm}

%%%%%%%%%%%%%%%%%%%%%%%%%%%%%%%%%%
\section{Results and Discussion}
\label{sec:benchmarks}
%%%%%%%%%%%%%%%%%%%%%%%%%%%%%%%%%%
All simulations were run on nodes with Intel Skylake CPU containing 40 cores and 96 GB of memory. The code was compiled with MPI 1.10.1, GCC 10.2.0, and HPX 1.4.1.
%%%%%%%%%%%%%%%%%%%%%%%%%%%%%%%%%%
\subsection{Validation of the Implementation}
\begin{figure}[tb]
    \centering
    \begin{tikzpicture}[thick,scale=0.77]
    \begin{axis}
    [
    grid=both,
    xlabel = Discretization parameter $h$,
    ylabel = Maximum relative error,
    title=\textbf{Plot for $h$ versus error}
    ]
    \addplot table [x=a, y=b, col sep=comma] {dh_linf.csv};
    \end{axis}
    \end{tikzpicture}
    \caption{Plot of the total error $e = \sum_{k} e^k$, see \eqref{eq:err} for different mesh sizes $h$ = 1 / $2^n$, where $n=2,3,..,6$. We expect the numerical error to decrease as the mesh size decreases.}
    \label{fig:ErrorPlot}
\end{figure}
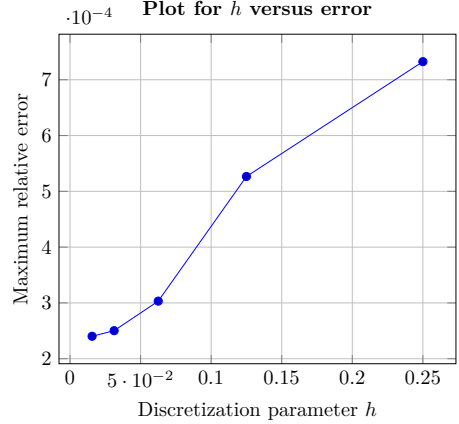

%%%%%%%%%%%%%%%%%%%%%%%%%%%%%%%%%%
We first validate the solver by considering a test setting and comparing the numerical solution with the analytical solution; we refer to Subsection~\ref{ss:exactsolution} for the analytical solution test details where the specific form of the external heat source $b$ is chosen so that the model has an analytical solution. We compute the error due to the numerical discretization following \eqref{eq:err}; the error depends on the timestep size and the mesh size and should decrease as the timestep size and mesh size decrease. From Figure~\ref{fig:ErrorPlot} we see that the numerical error is decreasing with a decrease in mesh size and this serve as a validation of the serial and \textit{distributed solver}. 

%%%%%%%%%%%%%%%%%%%%%%%%%%%%%%%%%%
\subsection{Shared Memory Implementation}
%%%%%%%%%%%%%%%%%%%%%%%%%%%%%%%%%%
We now study the speedup based on the instruction-level parallelism using multiple threads. The main idea is to distribute the $\SD$s uniformly among multiple threads located on the same computational node while sharing a common data structure. At every timestep, each thread is responsible for computing the temperature and updating the common data structure for the allocated $\SD$s. We study the speedup for 1, 2, and 4 CPU scenarios for fixed and variable problem sizes. The single CPU execution time is the baseline for the speedup plots.

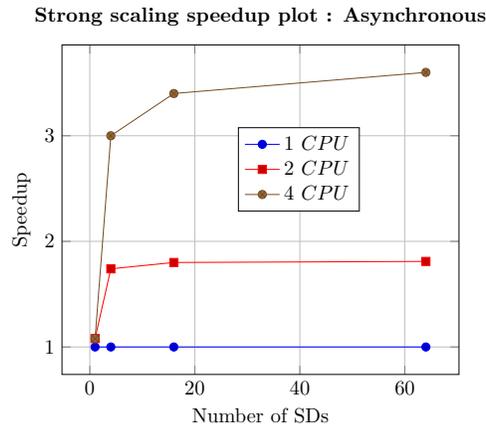
\begin{figure}[tb]
\begin{center}
\begin{tikzpicture}[thick,scale=0.77]
\begin{axis}[
  title = {\textbf{Strong scaling speedup plot : Asynchronous}},
  xlabel = {Number of $\SD$s},
  ylabel = {Speedup},
  grid = major,
  legend entries = {1 $CPU$, 2 $CPU$, 4 $CPU$},
  legend style={at={(0.75,0.75)}}]
  \addplot table [x=partitions, y=time, col sep=comma] {ss_async_speedup_1.csv};
  \addplot table [x=partitions, y=time, col sep=comma]{ss_async_speedup_2.csv};
  \addplot table [x=partitions, y=time, col sep=comma]{ss_async_speedup_4.csv};
\end{axis}
\end{tikzpicture}
\caption{Strong scaling results of the \textit{asynchronous solver} for mesh size = 400$\times$400 with $\epsilon=8h$ and no. of timesteps = 20 (\emph{i.e.} $N=20$ in $N\Delta t=T$). The entire mesh of size 400$\times$400 is divided into equal sized $\SD$s. The size of the $\SD$s is varied to keep the total mesh size constant, \emph{e.g.} For number of $\SD$s = 16, the partitioning is as follows: 4$\times$4 $\SD$s (\emph{i.e.} 4 $\SD$s along $X$ and $Y$ directions)  each of size 100$\times$100 (\emph{i.e.} 100 $\DP$s along $X$ and $Y$ directions in each $\SD$).}
\label{fig:SSAsyncSpeedup}
\end{center}
\end{figure}

\paragraph{\textbf{Strong scaling of asynchronous 2d nonlocal equation}}
In Figure~\ref{fig:SSAsyncSpeedup}, we present the scaling results for the asynchronous implementation of \eqref{eq:ForwardFiniteDiff} for the fixed problem size. We consider a fixed 400$\times$400 mesh and measure the effect of the decomposition into $\SD$s of different sizes. We consider $\SD$s of four sizes in four cases: 1) 1$\times$1 \emph{i.e.}, the entire square domain, 2) 2$\times$2 \emph{i.e.}, entire square domain is divided into a total of 4 partitions (2 along each of the axes), 3) 4$\times$4, and 4) 8$\times$8. The strong scaling plot in Figure~\ref{fig:SSAsyncSpeedup} indicates a linear dependence of the execution time on the number of CPUs.

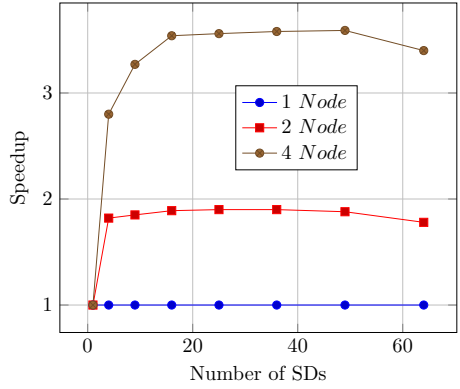
\begin{figure}[tb]
\begin{center}
\begin{tikzpicture}[thick,scale=0.77]
\begin{axis}[
  title = {\textbf{Weak scaling speedup plot : Asynchronous}},
  xlabel = {Number of $\SD$s},
  ylabel = {Speedup},
  grid = major,
  legend entries = {1 $Node$, 2 $Node$, 4 $Node$},
  legend style={at={(0.75,0.75)}}]
  \addplot table [x=partitions, y=time, col sep=comma] {ws_async_speedup_1.csv};
  \addplot table [x=partitions, y=time, col sep=comma]{ws_async_speedup_2.csv};
  \addplot table [x=partitions, y=time, col sep=comma]{ws_async_speedup_4.csv};
\end{axis}
\end{tikzpicture}
\caption{Weak scaling results of the \textit{asynchronous solver} for $\SD$ size = 50$\times$50 (\emph{i.e.} 50 $\DP$s along $X$ and $Y$ directions), $\epsilon=8h$ and no. of timesteps = 20 (\emph{i.e.} $N=20$ in $N\Delta t=T$). The number of $\SD$s is increased along both $X$ and $Y$ directions, keeping the size of the $\SD$s constant. The total mesh size is given by 50$n\times$50$n$, where $n$ is the number of $\SD$s.}
\label{fig:WSAsyncSpeedup}
\end{center}
\end{figure}

\paragraph{\textbf{Weak scaling of asynchronous 2d nonlocal equation}}
In Figure~\ref{fig:WSAsyncSpeedup}, we present the scaling of the asynchronous implementation of \eqref{eq:ForwardFiniteDiff} variable problem size.
We study the effect of increasing the mesh size by increasing the number of $\SD$s of fixed size 50$\times$50 along the $X$ and the $Y$ axes.
Eight different types of problem sizes that we considered are illustrated using the following examples: 1$\times$1 \emph{i.e.}, the total problem size is 50$\times$50; 2$\times$2 \emph{i.e.}, the total problem size is 100$\times$100; 3$\times$3 \emph{i.e.}, the total problem size is 150$\times$150; and 8$\times$8 \emph{i.e.}, the total problem size is 400$\times$400.
The weak scaling plot in Figure~\ref{fig:WSAsyncSpeedup} indicates a linear dependence of the execution time on the increase in problem size irrespective of the number of CPUs.

%%%%%%%%%%%%%%%%%%%%%%%%%%%%%%%%%%
\subsection{Distributed memory implementation}
%%%%%%%%%%%%%%%%%%%%%%%%%%%%%%%%%%
We now study the speedup based on data-level parallelism using multiple computational nodes. We distribute the $\SD$s uniformly across multiple computational nodes. Each computational node is responsible for computing the temperature corresponding to its $\SD$s. In this process, the computational nodes might exchange data to satisfy the nonlocal dependencies. In our experiments, we study the speedup for 1, 2, and 4 computational node scenarios for fixed and variable problem sizes. Single computational node execution time is the baseline for the speedup plots.

The distribution of $\SD$s across 1, 2 and 4 computational nodes is as follows:  1 Node: Entire square domain is on a single node; 2 nodes: Entire square domain is divided into 2 equal sized halves. For the number of $\SD$s = 4$\times$4, we divide the square domain into 2 halves of equal size -- 2$\times$4 and 2$\times$4. Each half is assigned to different computational nodes; and 4 nodes: Entire square domain is divided into 4 equal sized squares, each assigned to distinct computational nodes.

\begin{figure}[tb]
\begin{center}
\begin{tikzpicture}[scale=0.77]
\begin{axis}[
  title = {\textbf{Strong scaling speedup plot : Distributed}},
  xlabel = {Number of $\SD$s},
  ylabel = {Speedup},
  grid = major,
  legend entries = {1 $CPU$, 2 $CPU$, 4 $CPU$},
  legend style={at={(0.75,0.75)}}]
  \addplot table [x=partitions, y=time, col sep=comma] {ss_distributed_speedup_1.csv};
  \addplot table [x=partitions, y=time, col sep=comma]{ss_distributed_speedup_2.csv};
  \addplot table [x=partitions, y=time, col sep=comma]{ss_distributed_speedup_4.csv};
\end{axis}
\end{tikzpicture}
\caption{Strong scaling results of the \textit{distributed solver} for mesh size = 400$\times$400 with $\epsilon=8h$ and no. of timesteps = 20 (\emph{i.e.} $N=20$ in $N\Delta t=T$). The entire mesh of size 400$\times$400 is divided into equal sized $\SD$s. The size of the $\SD$s is varied to keep the total mesh size constant, \emph{e.g.} For number of $\SD$s = 16, the partitioning is as follows: 4$\times$4 $\SD$s (\emph{i.e.} 4 $\SD$s along $X$ and $Y$ directions)  each of size 100$\times$100 (\emph{i.e.} 100 $\DP$s along $X$ and $Y$ directions in each $\SD$).}
\label{fig:SSDistrSpeedup}
\end{center}
\end{figure}
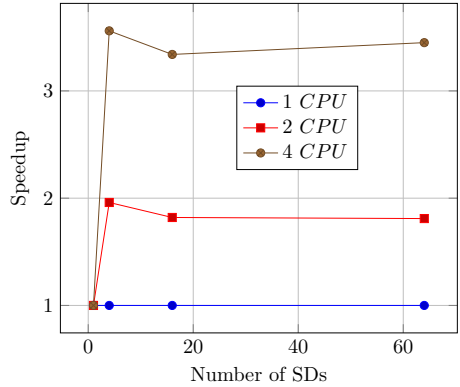

\paragraph{\textbf{Strong scaling of the distributed 2d nonlocal equation}}
In Figure~\ref{fig:SSDistrSpeedup}, we present the scaling of the distributed implementation of \eqref{eq:ForwardFiniteDiff} for a fixed problem size.
We study the effect of decomposition of a mesh with fixed size 400$\times$400 into a different number of $\SD$s to demonstrate the speedup. We consider $\SD$s of four sizes in four cases: 1) 1$\times$1 \emph{i.e.}, the entire square domain, 2) 2$\times$2 \emph{i.e.}, entire square domain is divided into a total of 4 partitions (2 along each of the axes), 3) 4$\times$4, and 4) 8$\times$8. The strong scaling plot in Figure~\ref{fig:SSDistrSpeedup} indicates a linear dependence of the speedup on the number of computational nodes.

\begin{figure}[tb]
\begin{center}
\begin{tikzpicture}[thick,scale=0.77]
\begin{axis}[
  title = {\textbf{Weak scaling speedup plot : Distributed}},
  xlabel = {Number of $\SD$s},
  ylabel = {Speedup},
  grid = major,
  legend entries = {1 $Node$, 2 $Node$, 4 $Node$},
  legend style={at={(0.75,0.75)}}]
  \addplot table [x=partitions, y=time, col sep=comma] {ws_distributed_speedup_1.csv};
  \addplot table [x=partitions, y=time, col sep=comma]{ws_distributed_speedup_2.csv};
  \addplot table [x=partitions, y=time, col sep=comma]{ws_distributed_speedup_4.csv};
\end{axis}
\end{tikzpicture}
\caption{Weak scaling results of the \textit{distributed solver} for $\SD$ size = 50$\times$50 (\emph{i.e.} 50 $\DP$s along $X$ and $Y$ directions), $\epsilon=8h$ and no. of timesteps = 20 (\emph{i.e.} $N=20$ in $N\Delta t=T$). The number of $\SD$s is increased along both $X$ and $Y$ directions, keeping the size of the $\SD$s constant. The total mesh size is given by 50$n\times$50$n$, where $n$ is the number of $\SD$s. The distribution of $\SD$s across the computational nodes is done using METIS library.}
\label{fig:WSDistrSpeedup}
\end{center}
\end{figure}
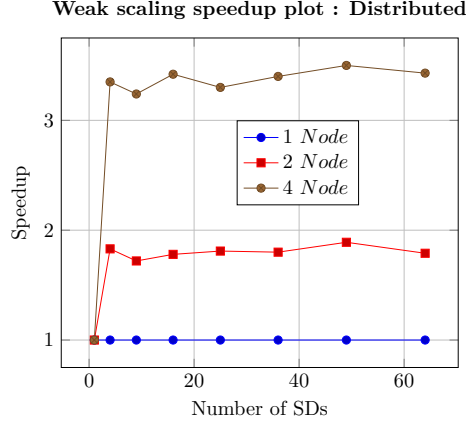

\paragraph{\textbf{Weak scaling of the distributed 2d nonlocal equation}}
In Figure~\ref{fig:WSDistrSpeedup}, we present the scaling of the distributed implementation of \eqref{eq:ForwardFiniteDiff} with variable problem size.
We study the effect of increasing the mesh size by increasing the number of $\SD$s where each $\SD$ is of fixed size 50$\times$50. 
%along the $X$ and the $Y$ axes.
Eight different types of problem sizes that we considered are illustrated using the following examples: 1$\times$1 \emph{i.e.}\ total problem size is 50$\times$50; 2$\times$2 \emph{i.e.}\ total problem size is 100$\times$100; 3$\times$3; 
%\emph{i.e.}\ total problem size is 150$\times$150; 
and 8$\times$8. 
%\emph{i.e.}\ total problem size is 400$\times$400. 
The weak scaling plot in Figure~\ref{fig:WSDistrSpeedup} indicates a linear dependence of the speedup with an increase in the number of computational nodes, irrespective of the problem size.

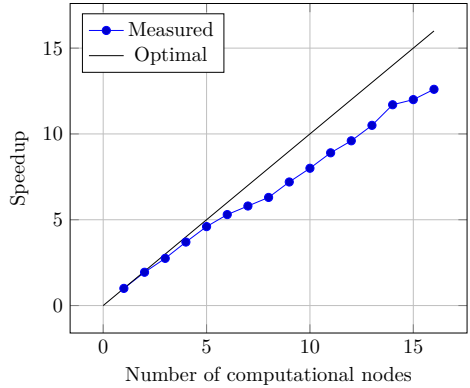
\begin{figure}[tb]
\begin{center}
\begin{tikzpicture}[thick,scale=0.77]
\begin{axis}
[
    grid=both,
    title = {\textbf{Distributed scaling: Domain decomposition using METIS}},
    xlabel = Number of computational nodes,
    ylabel = Speedup,
    legend entries = {Measured,Optimal},
    legend pos = north west
]
\addplot table [x=nodes, y=time, col sep=comma] {domain_decomposition_scaling_speedup.csv};
\addplot[domain=0:16] {x};
\end{axis}
\end{tikzpicture}
\caption{Distributed scaling results of the \textit{distributed solver} for mesh size = 800$\times$800, $\SD$ size = 50$\times$50 (\emph{i.e.} 50 $\DP$s along $X$ and $Y$ directions), $\epsilon=8h$ and no. of timesteps = 20 (\emph{i.e.} $N=20$ in $N\Delta t=T$). Total no. of $\SD$s is 16$\times$16 (\emph{i.e.} 16 $\SD$s along $X$ and $Y$ directions). The distribution of $\SD$s across the varying no. of computational nodes is done using METIS library.}
\label{fig:DDMetis}
\end{center}
\end{figure}

\paragraph{\textbf{Distributed scaling using METIS for mesh partitioning}}
We now study the scaling of the \textit{distributed solver} where we keep the size of the problem fixed (\emph{i.e} the mesh size is fixed) and increase the number of computational nodes; see the plot of speedup with different number of nodes in Figure~\ref{fig:DDMetis}. We consider a fixed 800$\times$800 mesh (800 grid points in $X$ and $Y$ directions); from this mesh we generate 16$\times$16 square $\SD$s with each $\SD$ consisting of 50$\times$50 grid points.  
%We divide the mesh of size 800$\times$800 into 16$\times$16 $\SD$s, each of size 50$\times$50. 
We then use METIS library for distribution of $\SD$s across multiple computational nodes.
Advantages of distributing the $\SD$s instead of the original fine mesh are as follows: the partitioning using METIS is very fast, since the number of $\SD$s is much smaller than the number of grid points in the original mesh; I/O time is reduced since we only need to read and write the $\SD$ allocation information; and, lastly, $\SD$s owned by a node can further be distributed across multiple threads within that node for parallel computation.

The distributed scaling plot in Figure~\ref{fig:DDMetis} indicates a linear relationship between the speedup and the number of computational nodes. As the number of computational nodes increases, the number of boundary $\SD$s requiring data exchange increases. This increase in the data exchange leads to a slight deviation from the straight line. 

\begin{figure}[tb]
\begin{minipage}[b]{0.47\linewidth}
\centering
\includegraphics[width=0.7\textwidth]{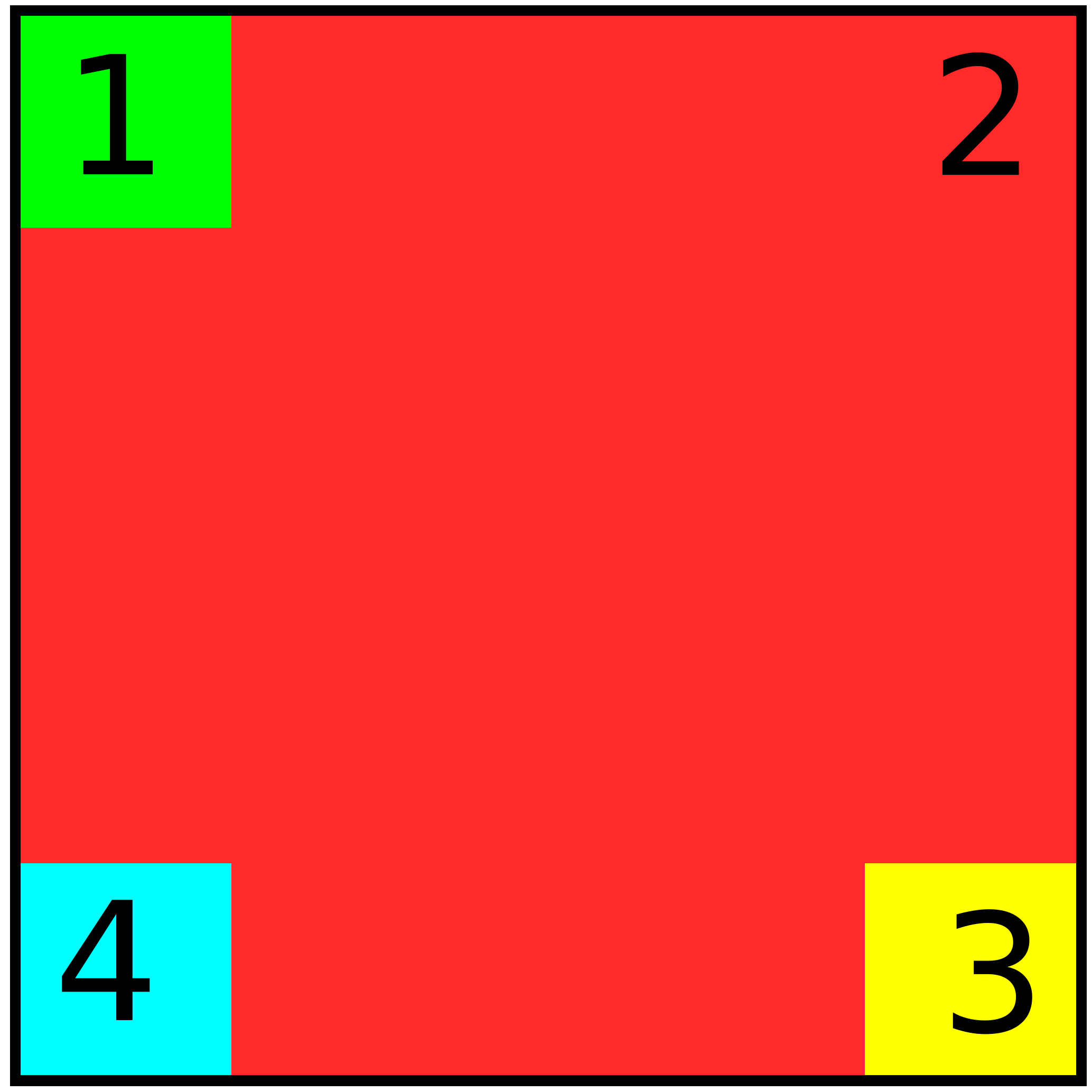}
\end{minipage}
\hspace{0.1cm}
\begin{minipage}[b]{0.485\linewidth}
\centering
\includegraphics[width=0.7\textwidth]{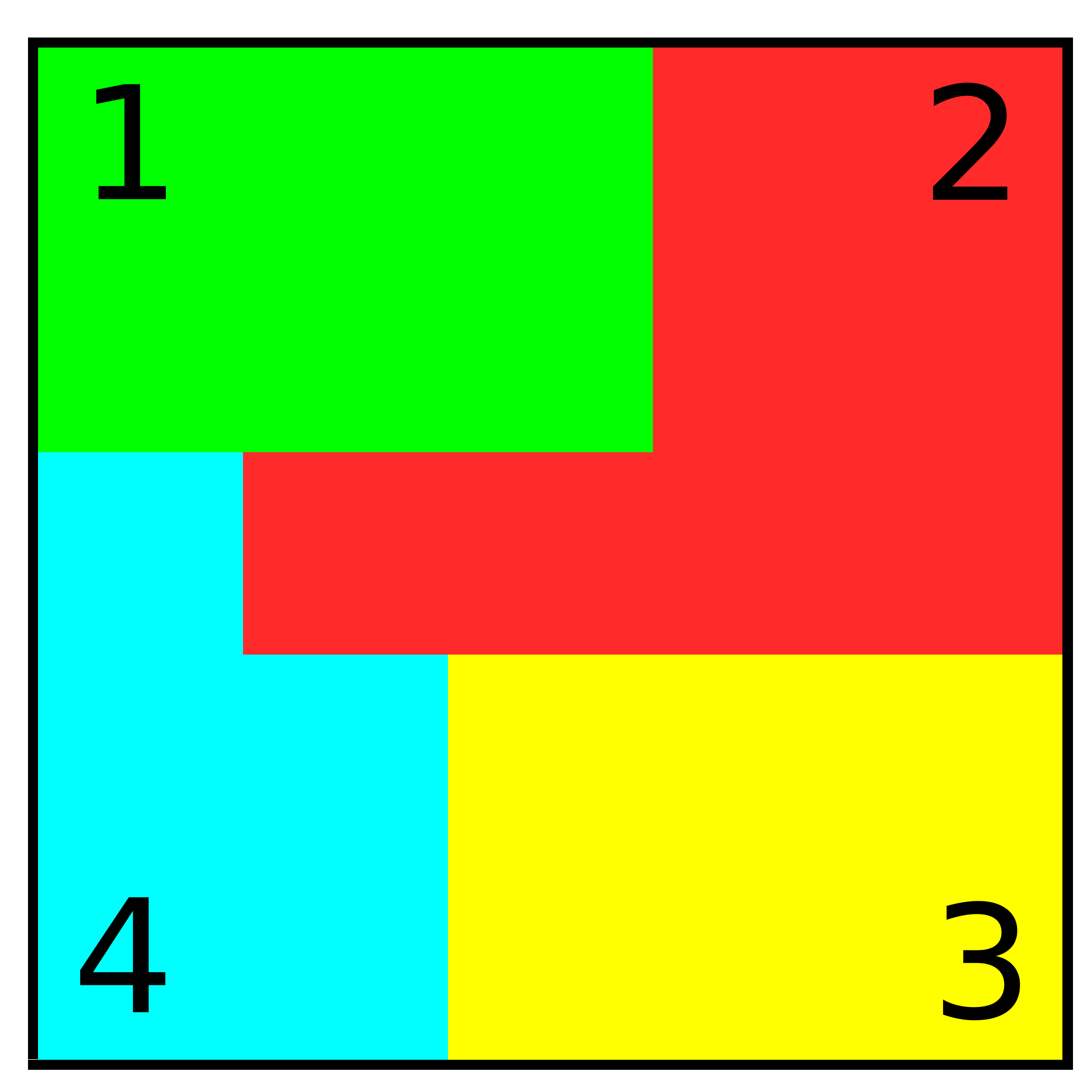}
\end{minipage}
\caption{Redistribution of 5$\times$5 $\SD$s across 4 computational nodes. Four colors denote the $\SD$s belonging to the four distinct nodes indicated by the numbers $1, 2, 3, 4$. Starting from a very imbalanced load distribution(left), within 3 iterations, the load balancing algorithm is able to achieve a very balanced distribution across 4 computational nodes(right).}
\label{fig:LoadBalancing}
\end{figure}

\paragraph{\textbf{Validation of the load balancing algorithm}}
To verify the load balancing algorithm in Algorithm~\ref{alg:LoadBalance}, we consider 4 symmetric nodes and assign a highly imbalanced load to each of them in the beginning. We demonstrate in Figure~\ref{fig:LoadBalancing} that within 3 iterations, the load balancing algorithm is able to redistribute the $\SD$s among various nodes with nearly balanced load distribution.

%%%%%%%%%%%%%%%%%%%%%%%%%%%%%%%%%%
\section{Conclusion and Future Work}
\label{s:conclusion}
%%%%%%%%%%%%%%%%%%%%%%%%%%%%%%%%%%
We proposed the ideas of -- (\textit{i}) coarsening the mesh for higher granularity, (\textit{ii)} efficient mesh partitioning using METIS, (\textit{iii}) hiding data exchange time using asynchronous computation and demonstrated good weak and strong scaling for the distributed implementation. We presented a novel load balancing algorithm by using HPX to schedule the tasks asynchronously (using \lstinline|hpx::future| and \lstinline|hpx::async|) and local control objects for synchronization. The major contribution is the novel load balancing algorithm that utilizes HPX's performance counters to address the specific challenges of nonlocal models. We used $\SD$s (sub-domains) as a unit of exchange to ensure simplicity in modelling the data exchange and to preserve the data locality. We proposed to preserve the contiguous locality obtained using the METIS mesh partitioning by borrowing the $\SD$s uniformly in all the spatial directions; the redistributed load after the load-balancing step minimizes the data exchange time. In one example, we could show that the proposed algorithm balanced a largely unbalanced domain within three iterations. For future work, we intend to investigate the addition of specific performance counters and networking counters. Larger node counts will be investigated which was not possible with our current allocation for this project. From the geometry perspective, a more complex non-square domains and three dimensions will be investigated. From the model perspective, a more complex models, \emph{e.g.} nonlocal mechanics~\cite{diehl2019review, Jha2020peri} will be investigated.

%%%%%%%%%%%%%%%%%%%%%%%%%%%%%%%%%%
\section*{Acknowledgements}
%%%%%%%%%%%%%%%%%%%%%%%%%%%%%%%%%%
We are grateful for the support of the Google Summer of Code program which funded P. Gadikar's summer internship.

%%%%%%%%%%%%%%%%%%%%%%%%%%%%%%%%%%
\section*{Supplementary materials}
%%%%%%%%%%%%%%%%%%%%%%%%%%%%%%%%%%
The source code to reproduce the results are available on GitHub\footnote{\url{https://github.com/nonlocalmodels/nonlocalheatequation}}.

%%%%%%%%%%%%%%%%%%%%%%%%%%%%%%%%%%%%%%%%%%
\section*{Copyright notice}
%%%%%%%%%%%%%%%%%%%%%%%%%%%%%%%%%%%%%%%%%%
\textcopyright 2021 IEEE. Personal use of this material is permitted. Permission from IEEE must be obtained for all other uses, in any current or future media, including reprinting/republishing this material for advertising or promotional purposes, creating new collective works, for resale or redistribution to servers or lists, or reuse of any copyrighted component of this work in other works.

\bibliographystyle{IEEEtran}
\bibliography{mybibfile}

\end{document}